\documentclass[11pt,a4paper]{article}
\usepackage[hyperref]{acl2019}
\usepackage{times}
\usepackage{enumitem}
\usepackage{latexsym}
\usepackage{graphicx}
\usepackage{hyperref}
\usepackage{url}
\usepackage{dblfloatfix}    % To enable figures at the bottom of page

\aclfinalcopy % Uncomment this line for the final submission
\title{A Multiscale Visualization of Attention in the Transformer Model}

\author{Jesse Vig \\
  Palo Alto Research Center \\
    3333 Coyote Hill Road \\
    Palo Alto, CA 94304 \\
  {\tt jesse.vig@parc.com}}

\date{}

\begin{document}
\maketitle
\begin{abstract}
The Transformer is a sequence model that forgoes traditional recurrent architectures in favor of a fully attention-based approach. Besides improving performance, an advantage of using attention is that it can also help to interpret a model by showing how the model assigns weight to different input elements. However, the multi-layer, multi-head attention mechanism in the Transformer model can be difficult to decipher. To make the model more accessible, we introduce an open-source tool that visualizes attention at multiple scales, each of which provides a unique perspective on the attention mechanism. We demonstrate the tool on BERT and OpenAI GPT-2 and present three example use cases: detecting model bias, locating relevant attention heads, and linking neurons to model behavior.
\end{abstract}

\section{Introduction}

In 2018, the BERT (Bidirectional Encoder Representations from Transformers) language representation model achieved state-of-the-art performance across NLP tasks ranging from sentiment analysis to question answering \citep{Bert2018}. Recently, the OpenAI GPT-2 (Generative Pretrained Transformer-2) model outperformed other models on several language modeling benchmarks in a zero-shot setting \cite{gpt2}.

Underlying BERT and GPT-2 is the Transformer model, which uses a fully attention-based approach in contrast to traditional sequence models based on recurrent architectures \cite{transformer_arxiv}.
An advantage of using attention is that it can help interpret a model by showing how the model assigns weight to different input elements \citep{align_translate, belinkov:2018:tacl}, although its value in explaining individual predictions may be limited \cite{attention_not_explanation}. Various tools have been developed to visualize attention in NLP models, ranging from attention-matrix heatmaps \citep{align_translate, neural_attention, rocktaschel2016reasoning} to bipartite graph representations \citep{visual_interrogation, Lee2017, seq2seqvisv1}. 

One challenge for visualizing attention in the Transformer is that it uses a multi-layer, multi-head attention mechanism, which produces different attention patterns for each layer and head. BERT-Large, for example, which has 24 layers and 16 heads, generates 24 $\times$ 16 = 384 unique attention structures for each input. \citet{JonesViz} designed a visualization tool specifically for multi-head attention, which visualizes attention over multiple heads in a layer by superimposing their attention patterns \cite{transformer_arxiv, tensor2tensor}.
 
In this paper, we extend the work of \citet{JonesViz} by visualizing attention in the Transformer at multiple scales. We introduce a high-level \textit{model view}, which visualizes all of the layers and attention heads in a single interface, and a low-level \textit{neuron view}, which shows how individual neurons interact to produce attention. We also adapt the tool from the original encoder-decoder implementation to the decoder-only GPT-2 model and the encoder-only BERT model.

\section{Visualization Tool}
We now present a multiscale visualization tool for the Transformer model, available at \url{https://github.com/jessevig/bertviz}. The tool comprises three views: an attention-head view, a model view, and a neuron view. Below, we describe these views and demonstrate them on the GPT-2 and BERT models. We also present three use cases: detecting model bias, locating relevant attention heads, and linking neurons to model behavior. A video demonstration of the tool can be found at \url{https://vimeo.com/340841955}. 

\begin{figure*}[h]
    \includegraphics[width=\linewidth]{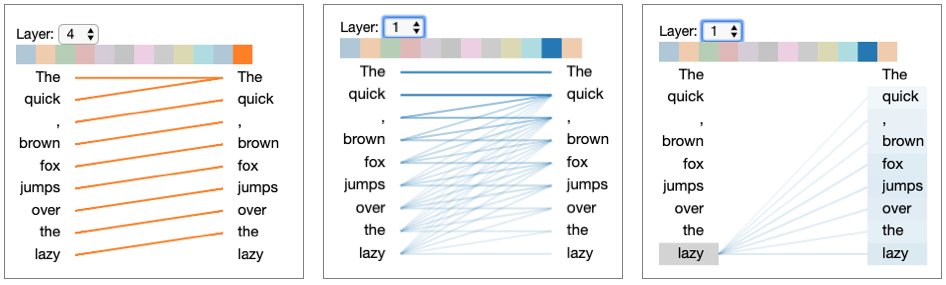}
    \vspace{-1.3em}
    \caption{Attention-head view for GPT-2, for the input text \textit{The quick, brown fox jumps over the lazy}. The left and center figures represent different layers / attention heads. The right figure depicts the same layer/head as the center figure, but with the token \textit{lazy} selected.}
    \vspace{1.3em}
    \label{fig:head_view_1_combined}
    \includegraphics[width=\linewidth]{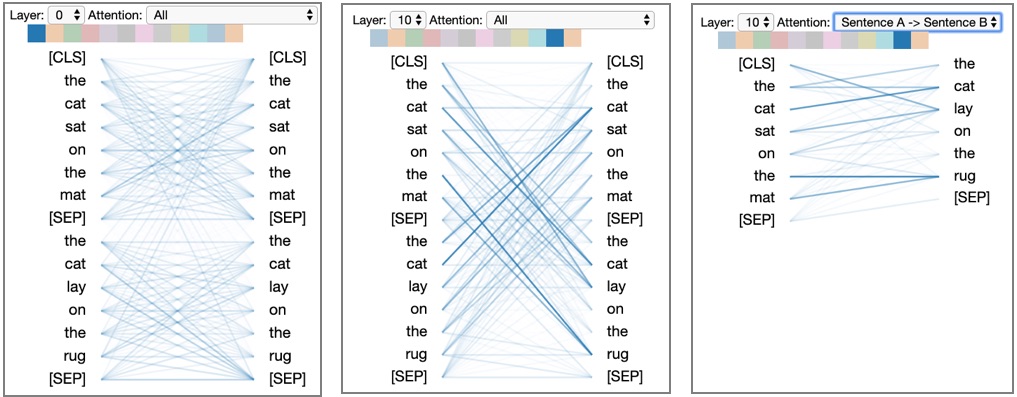}
    \vspace{-1.3em}
    \caption{Attention-head view for BERT, for inputs \textit{the cat sat on the mat} (Sentence A) and \textit{the cat lay on the rug} (Sentence B). The left and center figures represent different layers / attention heads. The right figure depicts the same layer/head as the center figure, but with \textit{Sentence A} $\rightarrow$ \textit{Sentence B} filter selected.}

    \label{fig:bert_heads}
    
\end{figure*}

\subsection{Attention-head view}

\begin{figure*}
\includegraphics[width=\linewidth]{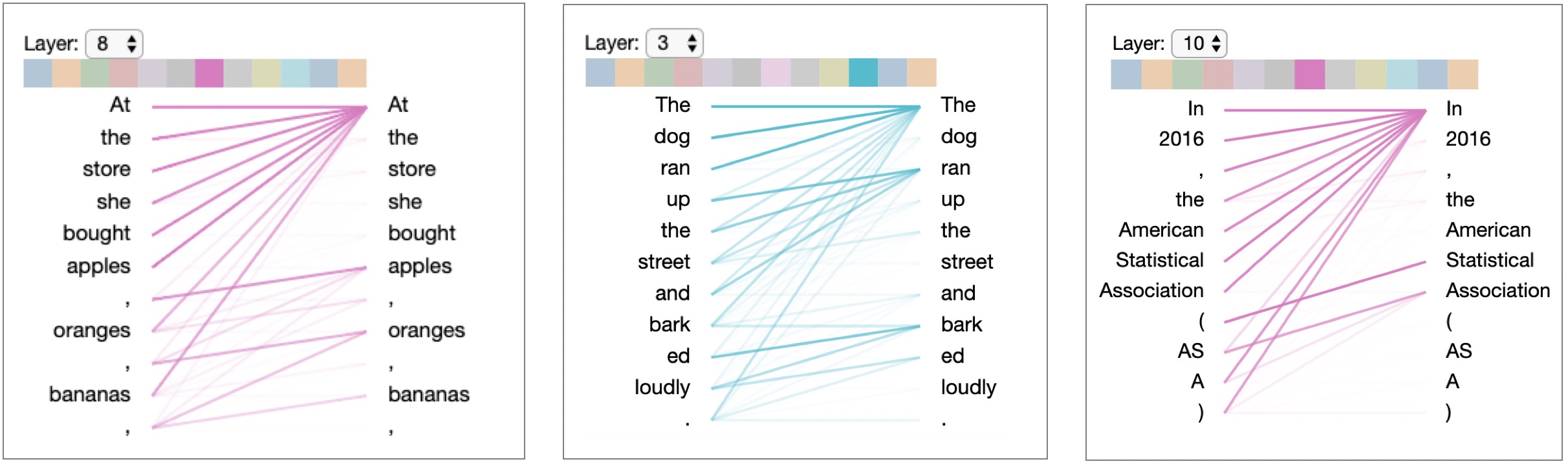}
\vspace{-1.7em}
    \caption{Examples of attention heads in GPT-2 that capture specific lexical patterns:  list items (left); verbs (center); and acronyms (right). Similar patterns were observed in these attention heads for other inputs. Attention directed toward first token is likely null attention \cite{vig-belinkov-structure-attn}.}
    \vspace{-.1em}
    \label{fig:example_combined}
\end{figure*}

\begin{figure*}[h]
\includegraphics[width=\linewidth]{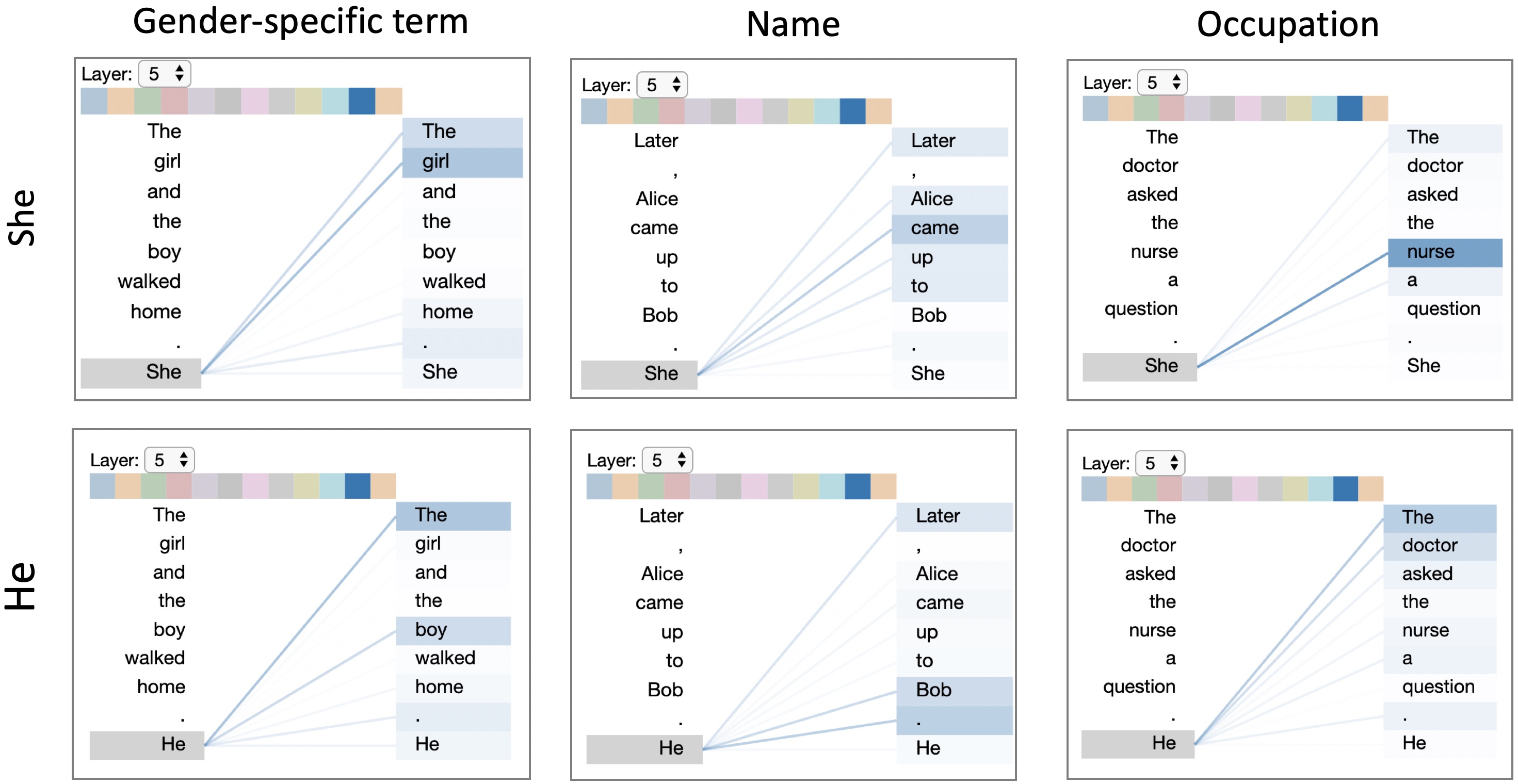}
\vspace{-1.7em}
\caption{Attention pattern in GPT-2 related to coreference resolution suggests the model may encode gender bias.} 
\label{fig:example_bias}
\end{figure*}

The \textit{attention-head view} visualizes the attention patterns produced by one or more attention heads in a given layer, as shown in Figure~\ref{fig:head_view_1_combined} (GPT-2\footnote{\textit{GPT-2 small} pretrained model.}) and Figure~\ref{fig:bert_heads} (BERT\footnote{\textit{BERT-base, uncased} pretrained model.}).  This view closely follows the original implementation of \citet{JonesViz}, but has been adapted from the original encoder-decoder implementation to the encoder-only BERT and decoder-only GPT-2 models. 

In this view, self-attention is represented as lines connecting the tokens that are attending (left) with the tokens being attended to (right). Colors identify the corresponding attention head(s), while line weight reflects the attention score. At the top of the screen, the user can select the layer and one or more attention heads (represented by the colored squares). Users may also filter attention by token, as shown in Figure~\ref{fig:head_view_1_combined} (right); in this case the target tokens are also highlighted and shaded based on attention weight. For BERT, which uses an explicit sentence-pair model, users may specify a sentence-level attention filter; for example, in Figure~\ref{fig:bert_heads} (right), the user has selected the \textit{Sentence A} $\rightarrow$ \textit{Sentence B} filter, which only shows attention from tokens in Sentence \textit{A} to tokens in \mbox{Sentence \textit{B}}.

Since the attention heads do not share parameters, each head learns a unique attention mechanism. In the head shown in Figure~\ref{fig:head_view_1_combined} (left), for example, each word attends to the previous word in the sentence. The head in Figure~\ref{fig:head_view_1_combined} (center), in contrast, generates attention that is dispersed roughly evenly across previous words in the sentence (excluding the first word). Figure~\ref{fig:bert_heads} shows attention heads for BERT that capture sentence-pair patterns, including a within-sentence pattern (left) and a between-sentence pattern (center).

Besides these coarse positional patterns, attention heads also capture specific lexical patterns, such as those as shown in Figure~\ref{fig:example_combined}. Other attention heads detected named entities (people, places, companies), paired punctuation (quotes, brackets, parentheses), subject-verb pairs, and other syntactic and semantic relations. Recent work shows that attention in the Transformer correlates with syntactic constructs such as dependency relations and part-of-speech tags \cite{raganato-tiedemann-2018-analysis, voita-etal-2019-pruning-heads, vig-belinkov-structure-attn}.

\vspace{5px}
\noindent {\bf Use Case: Detecting Model Bias}

\vspace{1px}
\noindent {One use case for the attention-head view is detecting bias in the model, which we illustrate for the case of conditional language generation using GPT-2. Consider the following continuations generated\footnote{Using GPT-2 small model with greedy decoding.} from two input prompts that are identical except for the gender of the pronouns (generated text underlined):}
\vspace{-.3em}
\begin{itemize}[leftmargin=*]
% \begin{itemize}
\item \textit{The doctor asked the nurse a question. She \textbf{\underline{said, ``I'm not sure what you're talking about.''}}}
\item \textit{The doctor asked the nurse a question. He \textbf{\underline{asked her if she ever had a heart attack.}}}
\end{itemize}

In the first example, the model generates a continuation that implies \textit{She} refers to \textit{nurse}.  In the second example, the model generates text that implies \textit{He} refers to \textit{doctor}. This suggests that the model's coreference mechanism may encode gender bias \cite{gender-bias-in-coreference, gender-bias-nlp}. Figure~\ref{fig:example_bias} shows an attention head that appears to perform coreference resolution based on the perceived gender of certain words. The two examples from above are shown in Figure~\ref{fig:example_bias} (right), which reveals that \textit{She} strongly attends to \textit{nurse}, while \textit{He} attends more to \textit{doctor}. By identifying a source of potential model bias, the tool could inform efforts to detect and control for this bias.

\subsection{Model View}

\begin{figure}[h]
    \includegraphics[width=.95\linewidth]{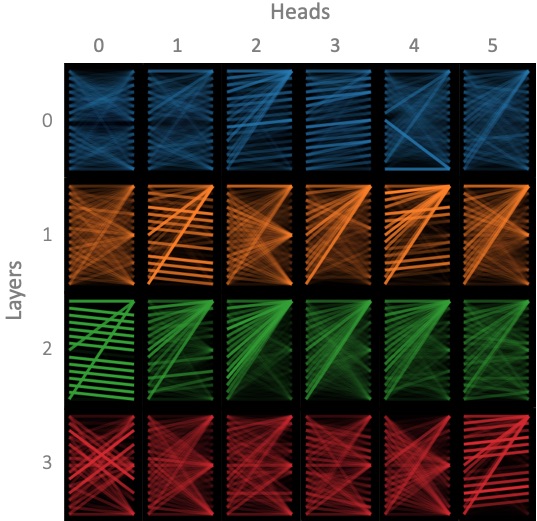}
    \caption{Model view of BERT, for same inputs as in Figure~\ref{fig:bert_heads}. 
    Excludes layers 4--11 and heads 6--11. }
    \label{fig:model_view}
    
\end{figure}

The \textit{model view} (Figure~\ref{fig:model_view}) provides a birds-eye view of attention across all of the model's layers and heads for a particular input. Attention heads are presented in tabular form, with rows representing layers and columns representing heads.  Each layer/head is visualized in a thumbnail form that conveys the coarse shape of the attention pattern, following the \textit{small multiples} design pattern \citep{Tufte1990}. Users may also click on any head to enlarge it and see the tokens.

\begin{figure*}[!t]
    \includegraphics[width=1\linewidth]{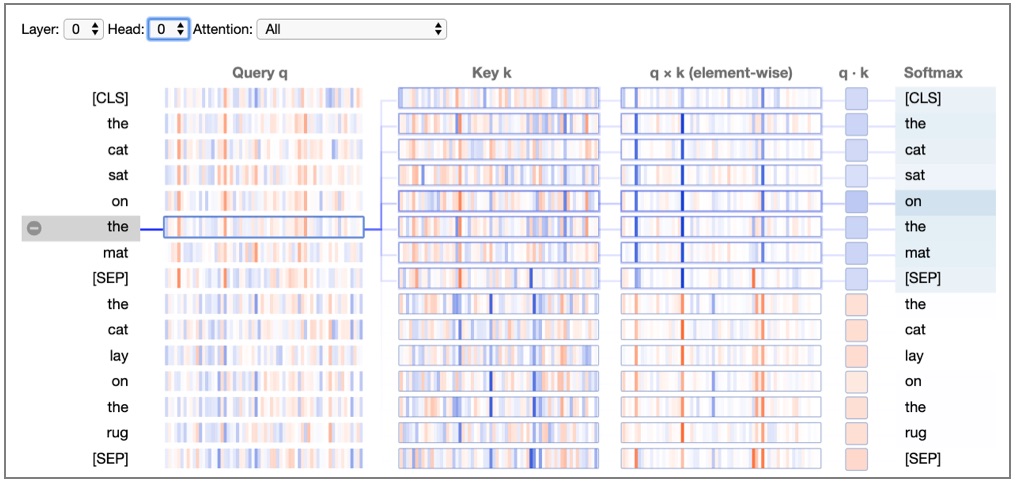}
    \vspace{-1.9em}
    \caption{Neuron view of BERT for layer 0, head 0 (same one depicted in Figure~\ref{fig:bert_heads}, left). Positive and negative values are colored blue and orange, respectively, with color saturation based on magnitude of the value. As with the attention-head view, connecting lines are weighted based on attention between the words. }
    \vspace{.9em}
    \label{fig:neuron_view}

    \includegraphics[width=1\linewidth]{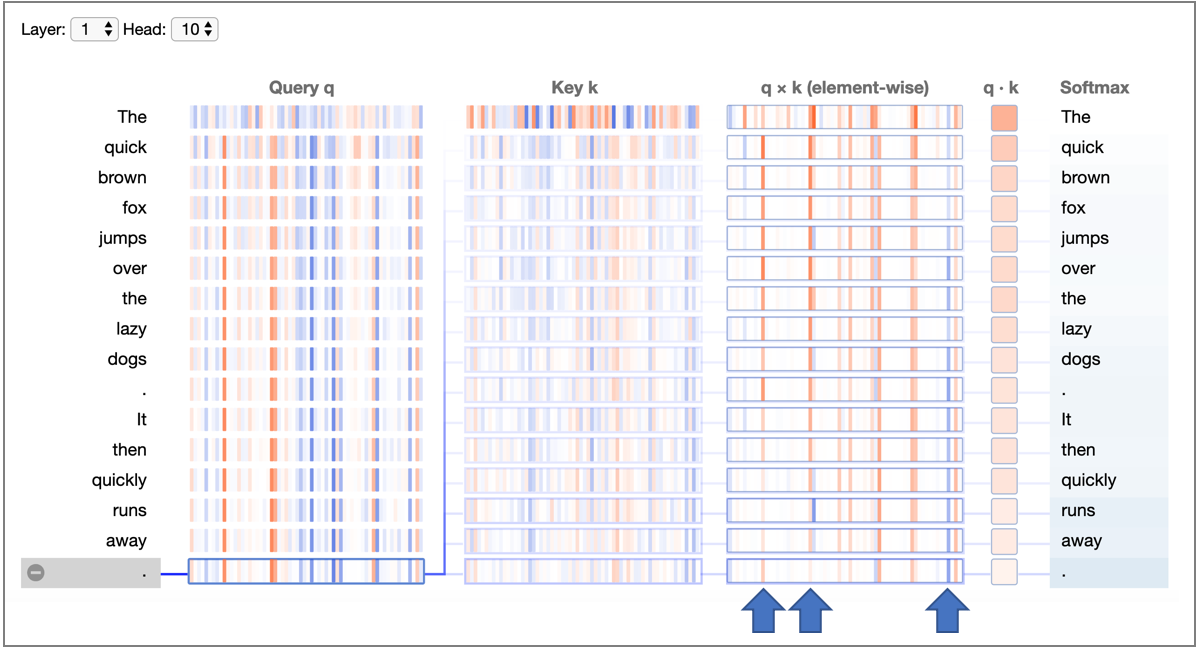}
    \vspace{-1.9em}
    \caption{Neuron view of GPT-2 for layer 1, head 10 (same one depicted in Figure~\ref{fig:head_view_1_combined}, center) with last token selected.  Blue arrows mark positions in the element-wise products where values decrease  with increasing distance from the source token (becoming darker orange or lighter blue).}
    \vspace{-.5em}
    \label{fig:neuron_view_2}
\end{figure*}

The model view enables users to quickly browse the attention heads across all layers and to see how attention patterns evolve throughout the model.

\vspace{5px}
\noindent {\bf Use Case: Locating Relevant Attention Heads}

\vspace{1px}
\noindent {As discussed earlier, attention heads in BERT exhibit a broad range of behaviors, and some may be more relevant for model interpretation than others depending on the task. Consider the case of paraphrase detection, which seeks to determine if two input texts have the same meaning. For this task, it may be useful to know which words the model finds similar (or different) between the two sentences. 
Attention heads that draw connections \textit{between} input sentences would thus be highly relevant.  The model view (Figure~\ref{fig:model_view}) makes it easy to find these inter-sentence patterns, which are recognizable by their cross-hatch shape (e.g., layer 3, head 0). These heads can be further explored by clicking on them or accessing the attention-head view, e.g., Figure~\ref{fig:bert_heads} (center). This use case is described in greater detail in \citet{vig-2019-bertviz}}.

\subsection{Neuron View}

The \textit{neuron view} (Figure~\ref{fig:neuron_view}) visualizes the individual neurons in the query and key vectors and shows how they interact to produce attention. Given a token selected by the user (left), this view traces the computation of attention from that token to the other tokens in the sequence (right).

Note that the Transformer uses scaled dot-product attention, where the attention distribution at position $i$ in a sequence $x$ is defined as follows:
\begin{equation}
\alpha_i = \mbox{softmax}\Big(\frac{q_i \cdot k_1}{\sqrt{d}}, \frac{q_i \cdot k_2}{\sqrt{d}},..., \frac{q_i \cdot k_N}{\sqrt{d}}\Big)
\end{equation}
where $q_i$ is the query vector at position $i$, $k_j$ is the key vector at position $j$, and $d$ is the dimension of $k$ and $q$. $N$=$i$ for GPT-2 and $N$=len($x$) for BERT.\footnote{GPT-2  only considers the context up to position $i$, while BERT considers the entire sequence.}
All values are specific to a particular layer / head.

The columns in the visualization are defined as follows:
\vspace{-.4em}
\begin{itemize}
    \setlength\itemsep{-.2em}
    \item \textbf{Query q}: The query vector of the selected token that is paying attention.
    \item \textbf{Key k}: The key vector of each token receiving attention. 
    \item \textbf{q$\times$k (element-wise)}: The element-wise product of the query vector and each key vector. This shows how individual neurons contribute to the dot product (sum of element-wise product) and hence attention.
    \item \textbf{q $\cdot$ k}: The dot product of the selected token's query vector and each key vector. 
    \item \textbf{Softmax}: The softmax of the scaled dot-product from previous column. This is the attention score.
\end{itemize}

\vspace{-.3em}
Whereas the attention-head view and the model view show \textit{what} attention patterns the model learns, the neuron view shows \textit{how} the model forms these patterns. 
For example, it can help identify neurons responsible for specific attention patterns, as discussed in the following use case.

\vspace{5px}
\noindent {\bf Use Case: Linking Neurons to Model Behavior}

\vspace{1px}
\noindent {To see how the neuron view might provide actionable insights, consider the attention head in Figure~\ref{fig:neuron_view_2}. For this head, the attention (rightmost column) decays with increasing distance from the source token. This pattern resembles a context window, but instead of having a fixed cutoff, the attention decays continuously with distance.}

The neuron view provides two key insights about this attention head. First, the attention weights appear to be largely independent of the content of the input text, based on the fact that all the query vectors have very similar values (except for the first token). Second, a small number of neuron positions (highlighted with blue arrows) appear to be mostly responsible for this distance-decaying attention pattern. At these neuron positions, the element-wise product $q \times k$ decreases as the distance from the source token increases (either becoming darker orange or lighter blue).

When specific neurons are linked to a tangible outcome, it presents an opportunity to intervene in the model \citep{Bau2019}. By altering the relevant neurons---or by modifying the model weights that determine these neuron values---one could control the attention decay rate, which might be useful when generating texts of varying complexity. For example, one might prefer a slower decay rate (longer context window) for a scientific text compared to a children's story. Other heads may afford different  types of interventions.

\section{Conclusion}
In this paper, we introduced a tool for visualizing attention in the Transformer at multiple scales. We demonstrated the tool on  GPT-2 and BERT, and we presented three use cases. For future work, we would like to develop a unified interface to navigate all three views within the tool.  We would also like to expose other components of the model, such as the value vectors and state activations. Finally, we would like to enable users to manipulate the model, either by modifying attention \citep{Lee2017, visual_interrogation, seq2seqvisv1} or editing individual neurons \citep{Bau2019}. 

\bibliography{main}

\begin{thebibliography}{21}
\expandafter\ifx\csname natexlab\endcsname\relax\def\natexlab#1{#1}\fi

\bibitem[{Bahdanau et~al.(2015)Bahdanau, Cho, and Bengio}]{align_translate}
Dzmitry Bahdanau, Kyunghyun Cho, and Yoshua Bengio. 2015.
\newblock \href {http://arxiv.org/abs/1409.0473} {Neural machine translation by
  jointly learning to align and translate}.
\newblock In \emph{Proc. {ICLR}}.

\bibitem[{Bau et~al.(2019)Bau, Belinkov, Sajjad, Durrani, Dalvi, and
  Glass}]{Bau2019}
Anthony Bau, Yonatan Belinkov, Hassan Sajjad, Nadir Durrani, Fahim Dalvi, and
  James Glass. 2019.
\newblock \href {https://arxiv.org/abs/1811.01157} {Identifying and controlling
  important neurons in neural machine translation.}
\newblock In \emph{{Proc. ICLR}}.

\bibitem[{Belinkov and Glass(2019)}]{belinkov:2018:tacl}
Yonatan Belinkov and James Glass. 2019.
\newblock \href {https://arxiv.org/abs/1812.08951} {Analysis methods in neural
  language processing: A survey}.
\newblock \emph{TACL}.

\bibitem[{Devlin et~al.(2018)Devlin, Chang, Lee, and Toutanova}]{Bert2018}
Jacob Devlin, Ming-Wei Chang, Kenton Lee, and Kristina Toutanova. 2018.
\newblock \href {https://arxiv.org/pdf/1810.04805.pdf} {{BERT: Pre-training of
  Deep Bidirectional Transformers for Language Understanding}}.
\newblock \emph{ArXiv Computation and Language}.

\bibitem[{Jain and Wallace(2019)}]{attention_not_explanation}
Sarthak Jain and Byron~C. Wallace. 2019.
\newblock \href {http://arxiv.org/abs/1902.10186} {Attention is not
  explanation}.
\newblock \emph{CoRR}, abs/1902.10186.

\bibitem[{Jones(2017)}]{JonesViz}
Llion Jones. 2017.
\newblock Tensor2tensor transformer visualization.
\newblock
  \url{https://github.com/tensorflow/tensor2tensor/tree/master/tensor2tensor/visualization}.

\bibitem[{Lee et~al.(2017)Lee, Shin, and Kim}]{Lee2017}
Jaesong Lee, Joong-Hwi Shin, and Jun-Seok Kim. 2017.
\newblock \href {https://doi.org/10.18653/v1/D17-2021} {Interactive
  visualization and manipulation of attention-based neural machine
  translation}.
\newblock In \emph{{EMNLP}: System Demonstrations}.

\bibitem[{Liu et~al.(2018)Liu, Li, Li, Srikumar, Pascucci, and
  Bremer}]{visual_interrogation}
Shusen Liu, Tao Li, Zhimin Li, Vivek Srikumar, Valerio Pascucci, and Peer-Timo
  Bremer. 2018.
\newblock \href {https://www.aclweb.org/anthology/D18-2007} {Visual
  interrogation of attention-based models for natural language inference and
  machine comprehension}.
\newblock In \emph{{EMNLP}: System Demonstrations}.

\bibitem[{Lu et~al.(2018)Lu, Mardziel, Wu, Amancharla, and
  Datta}]{gender-bias-nlp}
Kaiji Lu, Piotr Mardziel, Fangjing Wu, Preetam Amancharla, and Anupam Datta.
  2018.
\newblock \href {http://arxiv.org/abs/1807.11714} {Gender bias in neural
  natural language processing}.
\newblock \emph{CoRR}, abs/1807.11714.

\bibitem[{Radford et~al.(2019)Radford, Wu, Child, Luan, Amodei, and
  Sutskever}]{gpt2}
Alec Radford, Jeffrey Wu, Rewon Child, David Luan, Dario Amodei, and Ilya
  Sutskever. 2019.
\newblock \href
  {https://d4mucfpksywv.cloudfront.net/better-language-models/language_models_are_unsupervised_multitask_learners.pdf}
  {Language models are unsupervised multitask learners}.
\newblock Technical report.

\bibitem[{Raganato and Tiedemann(2018)}]{raganato-tiedemann-2018-analysis}
Alessandro Raganato and J{\"o}rg Tiedemann. 2018.
\newblock \href {https://www.aclweb.org/anthology/W18-5431} {An analysis of
  encoder representations in transformer-based machine translation}.
\newblock In \emph{EMNLP Workshop: BlackboxNLP}.

\bibitem[{Rockt{\"a}schel et~al.(2016)Rockt{\"a}schel, Grefenstette, Hermann,
  Kocisky, and Blunsom}]{rocktaschel2016reasoning}
Tim Rockt{\"a}schel, Edward Grefenstette, Karl~Moritz Hermann, Tomas Kocisky,
  and Phil Blunsom. 2016.
\newblock \href {https://arxiv.org/abs/1509.06664} {Reasoning about entailment
  with neural attention}.
\newblock In \emph{{Proc. ICLR}}.

\bibitem[{Rush et~al.(2015)Rush, Chopra, and Weston}]{neural_attention}
Alexander~M. Rush, Sumit Chopra, and Jason Weston. 2015.
\newblock \href {https://doi.org/10.18653/v1/D15-1044} {A neural attention
  model for abstractive sentence summarization}.
\newblock In \emph{{Proc. EMNLP}}.

\bibitem[{{Strobelt} et~al.(2018){Strobelt}, {Gehrmann}, {Behrisch}, {Perer},
  {Pfister}, and {Rush}}]{seq2seqvisv1}
H.~{Strobelt}, S.~{Gehrmann}, M.~{Behrisch}, A.~{Perer}, H.~{Pfister}, and
  A.~M. {Rush}. 2018.
\newblock \href {http://arxiv.org/abs/1804.09299v1} {{Seq2Seq-Vis: A Visual
  Debugging Tool for Sequence-to-Sequence Models}}.
\newblock \emph{ArXiv e-prints}.

\bibitem[{Tufte(1990)}]{Tufte1990}
Edward Tufte. 1990.
\newblock \emph{Envisioning Information}.
\newblock Graphics Press, Cheshire, CT, USA.

\bibitem[{Vaswani et~al.(2018)Vaswani, Bengio, Brevdo, Chollet, Gomez, Gouws,
  Jones, Kaiser, Kalchbrenner, Parmar, Sepassi, Shazeer, and
  Uszkoreit}]{tensor2tensor}
Ashish Vaswani, Samy Bengio, Eugene Brevdo, Francois Chollet, Aidan~N. Gomez,
  Stephan Gouws, Llion Jones, \L{}ukasz Kaiser, Nal Kalchbrenner, Niki Parmar,
  Ryan Sepassi, Noam Shazeer, and Jakob Uszkoreit. 2018.
\newblock \href {http://arxiv.org/abs/1803.07416} {Tensor2tensor for neural
  machine translation}.
\newblock \emph{CoRR}, abs/1803.07416.

\bibitem[{Vaswani et~al.(2017)Vaswani, Shazeer, Parmar, Uszkoreit, Jones,
  Gomez, Kaiser, and Polosukhin}]{transformer_arxiv}
Ashish Vaswani, Noam Shazeer, Niki Parmar, Jakob Uszkoreit, Llion Jones,
  Aidan~N Gomez, \L~ukasz Kaiser, and Illia Polosukhin. 2017.
\newblock \href {https://arxiv.org/pdf/1706.03762.pdf} {Attention is all you
  need}.
\newblock \emph{arXiv preprint arXiv:1706.03762}.

\bibitem[{Vig(2019)}]{vig-2019-bertviz}
Jesse Vig. 2019.
\newblock \href
  {https://debug-ml-iclr2019.github.io/cameraready/DebugML-19_paper_2.pdf}
  {{BertViz}: A tool for visualizing multi-head self-attention in the {BERT}
  model}.
\newblock In \emph{ICLR Workshop: Debugging Machine Learning Models}.

\bibitem[{Vig and Belinkov(2019)}]{vig-belinkov-structure-attn}
Jesse Vig and Yonatan Belinkov. 2019.
\newblock Analyzing the structure of attention in a transformer language model.
\newblock In \emph{ACL Workshop: BlackboxNLP}.

\bibitem[{Voita et~al.(2019)Voita, Talbot, Moiseev, Sennrich, and
  Titov}]{voita-etal-2019-pruning-heads}
Elena Voita, David Talbot, Fedor Moiseev, Rico Sennrich, and Ivan Titov. 2019.
\newblock \href {https://arxiv.org/abs/1905.09418} {Analyzing multi-head
  self-attention: Specialized heads do the heavy lifting, the rest can be
  pruned}.
\newblock \emph{arXiv preprint arXiv:1905.09418}.

\bibitem[{Zhao et~al.(2018)Zhao, Wang, Yatskar, Ordonez, and
  Chang}]{gender-bias-in-coreference}
Jieyu Zhao, Tianlu Wang, Mark Yatskar, Vicente Ordonez, and Kai-Wei Chang.
  2018.
\newblock \href {https://doi.org/10.18653/v1/N18-2003} {Gender bias in
  coreference resolution: Evaluation and debiasing methods}.
\newblock In \emph{NAACL-HLT}.

\end{thebibliography}
\bibliographystyle{acl_natbib}

\end{document}